\documentclass[prd,onecolumn]{revtex4}
\usepackage{graphicx}
\usepackage{epsfig}
\begin{document}
\newcommand{\be}{\begin{equation}}
\newcommand{\ee}{\end{equation}}
\newcommand{\ben}{\begin{eqnarray}}
\newcommand{\een}{\end{eqnarray}}
\newcommand{\bb}{\bibitem}
\newcommand{\ov}{\overline}
\newcommand{\wt}{\widetilde}
\newcommand{\nn}{\nonumber}




\title{Gravitational Effects of Varying Alpha Strings}

\author{J. Menezes\footnote{jmenezes@fc.up.pt}, P. P. Avelino\footnote{ppavelin@fc.up.pt}, and C. Santos\footnote{cssilva@fc.up.pt}}

\address{Centro de F\'{\i}sica do Porto e Departamento de F\'{\i}sica
da Faculdade de Ci\^encias da Universidade do Porto, Rua do Campo Alegre 687,
4169-007, Porto, Portugal}


\begin{abstract}

We study spatial variations of the fine-structure constant in the presence of static straight cosmic strings in
the weak-field approximation in Einstein gravity. We work in the context of a generic
Bekenstein-type model and 
consider a gauge kinetic function linear in the scalar field.  We determine 
an analytical form
for the scalar field and the string metric at large distances from the core. 
We show that the gravitational effects of $\alpha$-varying strings can be seen as a combination
of the gravitational effects of global and local strings. We also verify that at large distances to the core the
space-time metric is similar to that of a global string. We study the motion of test particles approaching 
from infinity and show that photons are scattered to infinity while massive particles are trapped in bounded
trajectories. 
We also calculate an overall limit on the magnitude of the variation of $\alpha$ 
for a GUT string, by considering suitable cosmological constraints coming from the Equivalence Principle.

\end{abstract}


\maketitle

\section{Introduction}

$ $\par

There has been a renewed interest in Cosmology in modeling and 
constraining possible space-time variations of the
so called $``$constants" of Nature. On the observational side, 
the analysis of quasar absorption spectra, by means of the many-multiplet 
method, reveals anomalies which could be interpreted in terms of a 
cosmological variation of $\alpha \equiv e^2/(4\pi\hbar\,c)$  
at low redshift \cite{Webb,Murphy}.
However, the situation is presently unclear since other more recent works 
also based on quasar absorption spectra seem to be consistent with no 
variation of $\alpha$ \cite{Quast,Srianand1,Srianand2}. 
Meanwhile other studies constrain variations of $\alpha$ at low redshift 
such as those based on atomic clocks \cite{Marion}, meteorites \cite{Olive} 
and the Oklo \cite{Fujii} natural nuclear reactor. At high redshifts 
variations of $\alpha$ are constrained by the Cosmic Microwave Background 
and Big Bang Nucleossynthesis \cite{hannestad,Avelino,Martins,Rocha,wmap1,sigurdson,wmap2}.

On the theoretical side, most work has been based on Bekenstein-type 
models \cite{Bekenstein} which provide a simple framework for studying 
$\alpha$ variability. Although a variation of $\alpha$ should in principle 
be accompanied by a change in the value of the others fundamental constants 
as well as the grand unification scale, one usually takes a phenomenological 
approach (see for example Ref.~\cite{OlivePos,Anchordoqui,Copeland,Nunes,Sandvik,vslcos,vslcos1,vs1,cosmic,linear,Lee,Parkinson,
Bennet,Nosso3,Barrow,Barrow2,Barrow3,Nosso,Magueijo,Nosso2}) 
neglecting possible variations of the other fundamental constants.

This paper is an extension of Ref.~\cite{Nosso} where the spatial variations
of $\alpha$ in the vicinity of static Abelian vortex in the 
context of Bekenstein-type models were investigated. There we considered a 
scalar field with a 
generic gauge kinetic function and computed the spatial variations of the 
fine-structure constant predicted in various models. We studied two 
special cases (a gauge kinetic function taking an exponential form which 
reproduces the original Bekenstein model \cite{Bekenstein}, and polynomial 
gauge kinetic function) and verified that, for certain classes of models, 
the classical Nielsen-Olesen vortex is still a valid solution. However, 
we showed that, in general, static strings solutions depart from 
the standard Nielsen-Olesen solution with the electromagnetic energy 
concentrated along the string core seeding the spatial variations of 
$\alpha$. We have also shown that Equivalence Principle constraints impose 
tight limits on the allowed variations of $\alpha$ induced by cosmic strings 
networks and calculated a conservative overall limit on the spatial variations 
of the fine-structure constant seeded by cosmic strings on cosmological 
scales, $\Delta\alpha/\alpha \lesssim 10^{-12}$, too small to have any 
significant cosmological impact. All these results were obtained in 
Minkowsky space-time. 

In the present paper, we generalize this previous study to include the 
gravitational field of the string. The paper is organized as follows. 
In section II we describe the Abelian-Higgs 
model coupled to Einstein gravity in the context of Bekenstein-type models and 
obtain field equations of motion. In section III we adopt an ansatz to describe 
static vortex solutions and simplify the field equations. We write the 
components of the energy-momentum tensor of the various components 
and discuss the boundary conditions. In section IV we describe the 
vortex solutions for the metric and for the scalar field that 
describes the variation of $\alpha$ far away from the string core. We analyse 
the non-conical nature of 
this metric and its similarity with that of a global string. In the following 
section we study the effect that $\alpha$ variability has in the geodesics of 
test particles. In section VI we estimate an overall limit on cosmological 
variations of the fine-structure constant seeded by these cosmic strings. 
We compare this result with that obtained in Ref.~\cite{Nosso}. 
Finally in section VII we summarize and discuss our results. Throughout this 
paper we shall use units in which $\hbar=c=1$ and the metric signature 
$+\,-\,-\,-$.

\section{The model}

$ $\par
We consider the Abelian Higgs model coupled to
Einstein gravity in a Bekenstein-type model \cite{Bekenstein}. Let us assume that the electric charge is a function of
the space-time coordinates, $e=e_0\,\epsilon(x^\mu)$ where $\epsilon$ is a real scalar field and $e_0$ is an arbitrary constant charge.
Let $\phi$ be a complex scalar field with a $U(1)$ gauge symmetry and $a_\mu$ be the gauge field. The action is given by
\ben
S\left[g_{\mu\nu},
\phi\,,\varphi\,,a_\mu,\phi^{\dagger}\right] &=& \int\,d^4\,x\,\sqrt{-g}\,(
-\frac{\mathcal{R}}{16\,\pi\,G}
+\left(D_\mu\,\phi\right)^{\dagger}\,\left(D^\mu\,\phi\right)-V(\phi,\phi^\dagger) \nn \\
&-& \frac{1}{4}B_F(\varphi)\,f_{\mu\nu}
\,f^{\mu\nu}+\frac{1}{2}\partial_\mu\,\varphi\,\partial^\mu\varphi\,), \label{action1}
\een
in which $B_F(\varphi)=\epsilon^{-2}(\varphi)$ is the gauge kinetic function and $\varphi$ is a massless scalar field. Here we assume that $V(\phi,\,\phi^\dagger)$ is the usual Mexican hat potential with
\be
V(\phi,\phi^\dagger) = \lambda\,(|\phi|^2-\eta^2)^2,
\ee
where $\lambda\,>\,0$ and $\eta$ is the symmetry breaking scale. 

In Eq.~(\ref{action1}),  $\mathcal{R}$ is the scalar curvature, $D_\mu \phi = (\nabla_\mu+i\,e_0\,a_\mu)\phi$ are
the gauge covariant derivatives, $\nabla_\mu$ is the covariant derivative and the electromagnetic field tensor is given by 
\be
f_{\mu\nu} = \partial_\mu\,a_\nu - \partial_\nu\,a_\mu.
\ee
The gauge kinetic function acts as the effective dielectric permittivity that can be phenomenologically assumed as an arbitrary function of $\varphi$. We take
\be
B_F(\varphi)=1+k\,\varphi
\ee
where we assume $k$ to be a constant. Equivalence Principle tests \cite{linear,Will,Damour} lead to the constraint $k \lesssim 5 \times 10^{-4} G^{1/2}$.

We consider a static straight self gravitating string. In order to determine 
its gravitational field outside the core we take the most general static 
axially symmetric metric \cite{Thorne} verifying the condition \cite{Synge} 
\be
(\delta^t_t + \delta^\theta_\theta)\,T + 2\,(T_t^t + T^\theta_\theta) = 0
\ee
where  $T_{\mu\nu}$ is the total energy-momentum tensor, 
$T$ is its trace and $\delta^\mu_\nu$ is the Delta Kronecker symbol.
This is given by \cite{Thorne}
\be
ds^2 = e^{\gamma}\,\left(dt^2-dr^2-dz^2\right)\,-{\tilde
\sigma}^2\,e^{-\gamma}\,d\theta^2\,, \label{conic}
\ee
where $\gamma$ and $\tilde{\sigma}$ are functions of $r$, the coordinate 
distance to the string core.
The Einstein equations then become
\ben
G_{\mu\nu} &=& \left(D_\mu\,\phi\right)^{\dagger}\,\left(D_\nu\,\phi\right)-\frac{1}{2}\,g_{\mu\nu}
\left(D_j\,\phi\right)^{\dagger}\,\left(D^j\,\phi\right)
+\frac{\lambda}{2}\,\left(|\phi|^2\,\,-\eta^2\right)^2
\,g_{\mu\nu} \nn \\
&+&\frac{1}{8}\,g_{\mu\nu}\,B_F(\varphi)\,f_{ab}\,f^{ab}
-\frac{1}{2}\,B_F(\varphi)\,f_{a\mu}\, f^{c\nu}\,g^{ac}
+\frac{1}{2}\partial_\mu\,\varphi\,\partial^\nu\varphi \nn \\
&-&\frac{1}{4}g_{\mu\nu}\partial_j\,\varphi\,\partial^j\,\varphi,
\label{einsteineq}
\een
and the equations of motion for the scalar and gauge fields  are
\ben
&&\nabla_\mu\,\nabla^\mu\,\varphi=\,-\frac{1}{4}\,\frac{\partial B_F(\varphi)}{\partial \varphi}\,f^2\label{varphieq}\\
&&D_\mu\,D^\mu\phi=-\frac{dV(\phi,\phi^\dagger)}{d\,\phi^\dagger}\label{phieq}\\
&&\nabla_\mu\,\left[B_F(\varphi)\,f^{\mu\nu}\right]\label{fmunueq}=j^\mu
\een
where the current $j^\mu$ is given by 
\be
j^\mu = ie_0 \left[\phi(D^\mu \phi)^\dagger -\phi^\dagger (D^\mu\phi)\right].
\ee

\section{The field equations}

$ $\par

We now look for static solutions choosing $``$vortex units$"$, i.e., in which the core size $(1/\sqrt{\lambda}\eta)$ is unity. Let us define the variables
\ben
R &=& \sqrt{\lambda}\,\eta\,r\\
\alpha &=& \sqrt{\lambda}\,\eta\,\tilde{\alpha}
\een 
to rewrite the equations of motion (\ref{einsteineq}-\ref{fmunueq}). 

A natural way to rewrite the fields that makes manifest the degrees of freedom of our model is
\ben
\phi(R) &=& \eta\,\chi(R)\,e^{i\,N\,\theta}\,, \label{Xansatz}\\
a_\theta(R) &=& \frac1{e_0\,R} \left[P(R) -N \right] \label{Pansatz}
\een
with all other components of $a_\mu$ set to zero. $\chi$ and $P$ are real functions of $R$ which are massive in the broken symmetry phase. For simplicity we choose $N=1$ in our ansatz.

The Einstein equations, Eq.~(\ref{einsteineq}), can be rewritten as 
\be
G_{\mu\nu} =\zeta\,M_{\mu\nu} +S_{\mu\nu}(\varphi),
\ee
where $\zeta$ is given by
\be
\zeta = 8\,\pi\,G\,\eta^2.
\ee
$M_{\mu\nu}$ and
$S_{\mu\nu}(\varphi)$ are the rescaled stress energy momentum tensors for
the vortex and scalar fields, whose elements are respectively given
by
\ben
&&M^t_t={\cal{E}}=e^{-\gamma}\,X^{\prime\,2}+e^{\gamma}\,\frac{X^2\,P^2}{\sigma^2}+\beta(\varphi)\,
\frac{P^{\prime\,2}}{\sigma^2}+(X^2-1)^2 \label{ttt}\\
&&M^R_R=-\mathcal{P}_R=-e^{-\gamma}\,X^{\prime\,2}+e^{\gamma}\,\frac{X^2\,P^2}{\sigma^2}-\beta(\varphi)\,
\frac{P^{\prime\,2}}{\sigma^2}+(X^2-1)^2\label{rrr}\\
&&M^\theta_\theta=-\mathcal{P}_\theta=e^{-\gamma}\,X^{\prime\,2}-e^{\gamma}\,\frac{X^2\,P^2}{\sigma^2}-\beta(\varphi)\,\frac{P^{\prime\,2}}{\sigma^2}+(X^2-1)^2\\
&&M^z_z=-\mathcal{P}_z=M^t_t
\een
where
\be
\beta(\varphi) = \frac{\lambda}{2\,e_0^2}\,\,B_F(\varphi)
\ee
is the Bogomoln'yi function and
\be
S^t_t=S^\theta_\theta=S^z_z=-S^r_r=\frac{1}{4}e^{-\gamma}\varphi^{\prime\,2}
\ee
with a prime meaning a derivative with respect to the
variable $R$.

The full equations of motion, Eq.~(\ref{einsteineq}-\ref{fmunueq}) are
\ben
&&\sigma^{\prime\prime}=-\zeta\,\sigma\,e^{\gamma}\left(\cal{E}-P_R\right) \label{ee1}\\
&&\left(\sigma\,\gamma^\prime\right)^\prime=\zeta\,\sigma\,e^\gamma\,
\left(P_R+P_\theta\right) \label{ee2}\\
&&\sigma\,^\prime\,\gamma^\prime=\sigma\,\frac{\gamma^{\prime\,2}}{4}+\zeta\,\sigma\,e^{\gamma}\,P_R
+\frac{\sigma}{4}\varphi^{\prime\,2} \label{ee3}\\
&&\left(\sigma\,\varphi^\prime\right)^\prime=\frac{\zeta}{2}\,\sigma\,e^{2\gamma}\,\frac{d\beta}{d\varphi}
\,\frac{P^{\prime\,2}}{\sigma^2} \label{ee4}\\
&&\frac{\left(\sigma\,X^\prime\right)^\prime}{\sigma}=\frac{P^2\,X\,e^{2\gamma}}{\sigma^2}+\,X(X^2-1)\,e^\gamma \label{ee5}\\
&&\left(\beta(\varphi)\,\frac{e^\gamma}{\sigma}\,P^\prime\right)^\prime=e^{2\gamma}\,\frac{X^2\,P}{\sigma} \label{ee6}
\een
with the Bianchi identity given by
\be	
\zeta
\mathcal{P}^\prime_R+\zeta\,
\left(\frac{\sigma^\prime}{\sigma}-\frac{\gamma^\prime}{2}\right)
(\mathcal{P}_R-\mathcal{P}_\theta)
+\zeta\,\gamma^\prime\,{\cal E} +
\zeta\gamma^\prime\,\mathcal{P}_R+\frac{\zeta}{4}\varphi^\prime\,k\,\left(\frac{P^\prime}{\sigma}\right)^2
= 0\,.
\ee
The Bogomoln'yi limit is when $B_F = 2\,e_0^2/\lambda$. In this limit the vortex equations, Eq.~(\ref{ee5}) and Eq.~(\ref{ee6}) can be reduced to a pair of first order differential equations.  

In the limit  $\zeta\,=\,0$ the string is not coupled to gravity. Using the results of Ref.~\cite{Nosso}, Eq.~(\ref{ee1}-\ref{ee6}) describe a vortex in a linearized Bekenstein theory when at the core and at very far distances from it one verifies the conditions:
\ben
\lim_{R \rightarrow 0}\,\chi(R) = 0,\,\,\,\,\,\,\,\lim_{R \rightarrow \infty}\,\chi(R) = 1,\\
\lim_{R \rightarrow 0}\,P(R) = 1,\,\,\,\,\,\,\,\lim_{R \rightarrow \infty}\,P(R) = 0,\\
\lim_{R \rightarrow 0}\,B_F(R) = 1,\,\,\,\,\,\,\,\lim_{R \rightarrow 0}\,\frac{d\,\varphi}{dR} = 0.
\een

In this limit which there are no variation of $\alpha$, which by Ref.~\cite{Nosso} means that in this case the vortex must be the Nielsen-Olesen one. Thus, the vortex field rapidly falls off to their vacuum value outside the core.

For $\zeta \ll 1$ we have a weak gravitational coupling regime, that is, such the one at Grand Unified Scales ($\eta \simeq 10^{16}$ Gev, i.e., $\zeta\,\sim\,10^{-6}$). In this case the gravitational corrections are weak enough and we can still assume that the vortex fields do quickly asymptote the vacuum outside the core.

\section{Asymptotic Solutions}

$ $\par

Up to order $O(\zeta^2)$ one gets for the asymptotic vortex geometry and the scalar field
\ben
&&\sigma=(1-A)R+B,\label{alphasol}\\
&&\gamma=D,\\
&&\varphi=
k\,H\,\ln\left((1-A)R+B\right),\label{varphisol}
\een
which is a Levi-Civita \cite{Civita} solution for the
metric with $A$, $B$, $D$ and $H$ given by
\ben
A&=& \zeta \int_0^{\infty} R\, \left({\cal E}_0-{\cal P}_{0R} \right)\,dR,  \label{A} \\
B&=& \zeta \int_0^{\infty} R^2\, \left({\cal E}_0-{\cal P}_{0R} \right) \,dR, \label{B} \\
D&=& \zeta \int_0^{\infty} R\, {\cal P}_{0R}\,dR,   \label{D} \\
H&=& \zeta \int_0^{\infty} \, \frac{P_0^{\prime\,2}}{2\,R}\,dR. \label{H}
\een
Here ${\cal E}_0$ and ${\cal P}_{0R}$ are the rescaled energy density and radial pressure of the Nielsen-Olesen string which are given by Eq.~(\ref{ttt}) and Eq.~(\ref{rrr}), with $\gamma=0$, $\sigma=R$, $B_F=1$ and $\chi_0$ and $P_0$ are the Nielsen-Olesen vortex fields. Note
that since these fields rapidly fall off to their value
vacuum outside the core, the integrals $A$, $B$, $D$ and $H$ rapidly converge( to their
asymptotic constant values). 

Let us define
\be
\hat t=t\,e^{\frac{D}{2}},\,\,\,\,\,\hat z=z\,e^{\frac{D}{2}},\,\,\,\,\,\hat R=R\,e^{\frac{D}{2}}. \label{deff}
\ee
One gets for the asymptotic form of the metric
\be
\label{assympmetric} ds^2 = \hat{R}^{k^2\,H^2}\,\left(d{\hat t}^2
-d{\hat z}^2-d{\hat R}^2\right) -(1-4\,G\,\mu)^2\,\hat
R^{2-k^2\,H^2}\,d\theta^2
\ee
where $\mu$ is the string energy per unit length, given by $4\,G\,\mu = A + D$, and
\be
\varphi \simeq \ln\,(R)^{k\,H}. \label{var}
\ee

We note that for $k\,H\,\ll 1$ e $R \gg 1$,
\be
R^{k^2\,H^2} \simeq 1\,+\,k^2\,H^2\,\ln R,
\ee
and therefore the massless scalar field introduces an asymptotic form
for the metric reminiscent of the global string with a conical deficit
angle \cite{Harari,Cohen,Vilenki}. 



The non-conical effects become important when 
$\hat{R}^{k^2\,H^2}\,\approx\,e^2$, that is,
\be
\hat{R}\,\approx\,e^{\frac{2}{k^2\,H^2}}\,>\,e^{(3 \times 10^{26})} \label{rtil}
\ee
where we used
\be
H = \zeta\,h
\ee
which for a GUT string gives numerically that $h\,\simeq 10^{-1}$  and thus $H\,\simeq\,10^{-7}$.
For a typical core size $10^{-32}\,m$ one gets in vortex units that $k < 8 \times 10^{-7}$ which means non-conical effects
for scales 
\be
r > e^{3 \times 10^{26}} \times 10^{-32}\,m \simeq 10^{10^{26}}\,Mpc,
\ee
i.e., well beyond any reasonable cosmological scale.

We can also verify that the asymptotic form for the metric in Eq.~(\ref{assympmetric}) is equivalent to that of a dilatonic string in the Einstein
frame \cite{Ruth} with $a \neq -1$ where $e^{2a\psi}$ is the coupling of the vortex fields to the dilaton $\psi$.

\section{Geodesics}

$ $\par

In this section we discuss the motion of test particles on the presence of varying $\alpha$ strings.
These follow the
geodesics in the space-time presented in Eq.~(\ref{assympmetric}).

The radial motion of a test particle in a plane transverse to the
string, $d\hat z=0$, is given by 
\be 
\dot{\hat R}^2 +
\frac{J^2}{{\hat R}^2} + \frac{p}{{\hat R}^\nu}=\frac{E^2}{\hat{R}^{2\nu}} \label{geod} 
\ee 
where $\nu = (k\,H)^2$ and the dot denotes
a derivative with respect to the proper time along a time-like
geodesic, or an affine parameter for photons. The parameter $p$ is
either one or zero, representing either a massive particle or
photon respectively. $E$ and $J$ are constants of the motion
representing the particle energy and angular momentum respectively, and are
given by
\ben
&&E=g_{{\hat t \hat t}} {\dot {\hat t}}=\hat{R}^{\nu}\,{\dot {\hat t}}\\
&&J=\frac{g_{\theta\theta}\,{\dot\theta}}{1-4\,G\,\mu}=
-(1-4\,G\,\mu)\,\hat R^{ 2-\nu}\,\dot\theta
\een

It is useful to redefine the radial coordinate $\rho$ by using:
\be
\rho =(\hat{R}^{\nu+1})/(\nu+1). \label{roh}
\ee
Substituting Eq.~(\ref{roh}) in Eq.~(\ref{geod}),
one gets the $\rho$-radial motion as that of a unit mass particle of energy $E^2/2$
\be
\frac{\dot{\rho}^2}{2} + \frac{J^2}{2}\,\left[\rho\,(\nu+1) \right]^{-\frac{2\,(\nu-1)}{\nu+1}}
+\frac{p}{2}\,\left[\rho(\nu+1) \right]^{\frac{\nu}{\nu+1}}=\frac{E^2}{2}
\ee
with an effective potential given by 
\be
\label{3upotential}
U_{_{\rm eff}}(\rho,\,\nu) =
\frac{J^2}{2}\left[\rho(\nu+1)\right]^{-\frac{2(\nu-1)}{\nu+1}}
+\frac{p}{2}\left[\rho(\nu+1)\right]^{\frac{\nu}{\nu+1}} 
\ee 

\begin{figure}
\begin{center}
\includegraphics*[width=6cm]{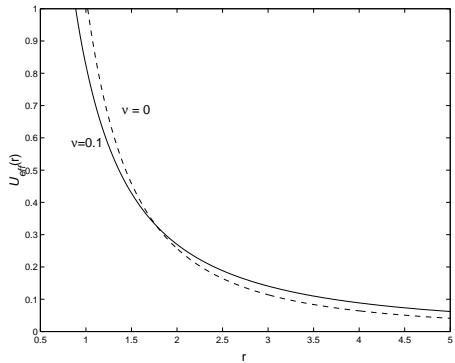}
\end{center}
\caption{
$U_{eff}$ as a function of $\rho$ for a photon with $J^2\,=\,2$.
The solid lines are for $\nu = 0.4$, $\nu = 0.5$ and $\nu = 0.6$, while the dashed line is for $\nu = 0$, in which $\alpha$ does not vary.}
\label{photon}
\end{figure}

Since $\nu$ is of order $O(\zeta^2)$,
to leading order one gets
\be
U_{_{\rm eff}}(\rho,\,\nu)
\simeq \frac{J^2}{2}\left[\rho^{-2(1-2\nu)}\,(1-2\nu)\right]
+\frac{p}{2}\left[\rho^{\nu}\right]  \label{potefe}
\ee

\begin{figure}
\begin{center}
\includegraphics*[width=6cm]{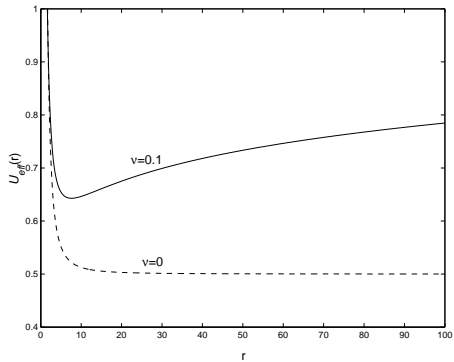}
\end{center}
\caption{
 $U_{eff}$ as a function of $\rho$ for a massive particle with $J^2\,=\,2$.
The dashed line is for $\nu=0$, in which $\alpha$ does not vary. The solid lines are for $\nu=0.4$, $\nu = 0.5$, $\nu = 0.6$ and $\nu = 0.7$. For $\nu=0.4$, $U_ {eff}$
has a minimum at $\rho=0.32$ while for $\nu=0.6$ the minimum is at $\rho = 1.3\times 10^{-3}$. Meanwhile for $\nu=0.7$ the effective potential has a maximum at $\rho=2.45$.}
\label{mass}
\end{figure}

If one considers $k=0$ in Eq.~(\ref{assympmetric}) one gets the asymptotic space-time of a local string in the weak coupling regime which is given by \cite{Vilenki}
\be
d\,s^2 = d\,\hat{t}^2 - d\,\hat{z}^2 - d\,\hat{R}^2 - (1-4\,G\,\mu)^2\,\hat{R}^2\,d\,\theta^2. \label{metr}
\ee
Thus the effective potential is a scattering
potential not only for photons but also for massive particles as it is shown for $J^2=2$ and $\nu=0$ in Fig.~\ref{photon} and Fig.~\ref{mass} (dashed lines).

For $k\,\neq\,0$(that is $\nu\,\neq\,0$), in Eq.~(\ref{potefe}), there are three
different possible asymptotic regimes for $U_{eff}$ depending whether $\nu\,<\,0.5$, or  $\nu\,=\,0.5$ or $\nu\,>\,0.5$. 
We notice that for GUT strings $\nu\,<\,0.5$. In this case, $U_{eff}$ is still a scattering potential for 
photons but now the scattering occurs asymptotically at closer distances 
to the string, as one can check for $\nu=0.4$ through Fig.~\ref{photon} (solid line). Nevertheless 
for massive particles the trajectories are bounded. The 
effective potential has a minimum given by 
\be
\rho^{*} = \left[ 8\,J^2\,\frac{(\nu-1/2)^2}{\nu}\right]^{\frac{1}{2-3\,\nu}}
\ee
and it is plotted in Fig.~\ref{mass} (solid line), for
$\nu\,=\,0.4$. 

For $\nu=0.5$ and for photons the effective potential is constant($U_{eff}=J^2/2$). Thus for $E\,>\,J$ photons behave like free particles, while for massive particles all the trajectories are bounded as one can check through Fig.~\ref{photon} and Fig.~\ref{mass}(solid lines).

Finally for $\nu > 0.5$ photons still behave like free particles for $E >J$. Meanwhile for massive particles the energy potential has an extremum at
\be
\bar{\rho} = \left[\frac1{J^2}\,\frac{\nu}{2(1-2\,\nu)^2} \right]^{\frac1{3\nu-2}}
\ee
that for $0.5<\nu<2/3$ is a minimum while for $\nu>2/3$ is a maximum. Thus the trajectories of massive particles are bounded for $0.5 < \nu < 2/3$ the particles are scattered by the potential as one can see in Fig.~\ref{mass} and Fig.~\ref{photon}(solid lines).
These geodesics are similar to those for a global string, which present bounded trajectories
for massive particles. In fact the effective potential in  Eq.~(\ref{potefe}) can be rewritten as
\be
U_{eff}(\rho,\nu)\,=\,U_{eff}(\rho,\nu=0)\,\rho^{4\,\nu}\,(1-2\,\nu)-\frac{p}{2}\,\rho^{4\nu}(1-2\,\nu)+\frac{p}{2}\,\rho^{\nu},
\ee
i.e., a contribution coming from the local nature of the string space-time corresponding to $U_{eff}(\rho,\nu=0)$, combined
with terms of global nature. The functions $U_{eff}(\rho,\nu=0)$ and $U_{eff}(\rho,\nu)$ always intercept in a single
point for a certain distance to the core.

We note that for a GUT string $\nu$ is less than $10^{-26}$. This means that in Fig.~\ref{photon} and Fig.~\ref{mass}
we should be looking to much larger scales so that in fact the changes in the geodesics are not mensurable. 

\section{Constraints on Variations of $\alpha$}

$ $\par

We now estimate an upper limit for the spatial variations of $\alpha$. Given that   
\be
\varphi = \ln\left(\frac{e}{e_0}\right) 
\ee
where $e_0$ is an arbitrary constant electron charge, we get from Eq.~(\ref{var}) that  
\be
\alpha = \frac{e_0^2}{4 \pi}\,R^{2 k H}. \label{alf}
\ee
Then the fine-structure increases with the distance to the core of the string. Taking into account that we cannot observe scales
larger than the horizon($\sim 10^{4}$ Mpc) and that is unlikely to probe variations of $\alpha$ at a distance much smaller than $1pc$ from a cosmic string, Eq.~(\ref{alf}) implies an overall limit on observable variations of the fine-structure constant seeded by GUT cosmic strings of
\be
\frac{\Delta\,\alpha}{\alpha}\, \lesssim 10^{-12}.
\ee
This limit is too small to have any significant cosmological impact. These variations are consistent with the variation of $\alpha$ obtained in Ref.~\cite{Nosso} which neglected gravitational effects, because the gravitational perturbations to the Minkowsky metric are small for a GUT string.

\section{Conclusions}

$ $\par

In this article we studied string vortex solutions in Bekenstein-type models
with a gauge kinetic 
function linear in the scalar field. The spatial variations of $\alpha$ produce an asymptotic form for
the string metric which is a reminiscent of that of global strings. We showed that for GUT strings the non-conical effects
in the metric become evident only for very large scales (well beyond the present Hubble radius). We have shown that photons approaching the 
string from infinite distances to the core will be scattered while massive particles are trapped in bounded trajectories. For GUT strings
one gets $\Delta\,\alpha/\alpha\,\lesssim\, 10^{-12}$ in agreement with the result presented in Ref.~\cite{Nosso} which is too small to be of any cosmological relevance.
We shall leave for future work the study of dilatonic $\alpha$-varying strings. However, we anticipate that these should be similar to strings in axion-dilatonic gravity \cite{axion} with
a generic kinetic gauge coupling. In this case it should be possible to relate the variations of fine-structure constant with those of the gravitational constant.

\begin{acknowledgments}
This work was funded by FCT (Portugal), through grant POCTI/CTE-AST/60808/2004, in the framework of the POCI2010 program, supported by FEDER.
J. Menezes was supported by a Brazilian Government grant - CAPES-BRAZIL 
(BEX-1970/02-0). 
\end{acknowledgments}

\end{document}